\documentclass{aa} 
\usepackage{txfonts} 
\usepackage{natbib}	 
\usepackage{ulem} 
\usepackage{txfonts,epsfig,graphicx,url,twoopt} 
\usepackage[breaklinks=true]{hyperref} 
\usepackage{dblfloatfix} 
\usepackage{float}
\usepackage{capt-of} 
\newcommand{\RNum}[1]{\uppercase\expandafter{\romannumeral #1\relax}}
\usepackage{multirow}
\hypersetup{colorlinks=true,citecolor=blue} 
\bibpunct{(}{)}{;}{a}{}{,} 
\begin{document} 
\title{Resolving the distance controversy for Sharpless 269:}
\subtitle{A possible kink in the outer arm}
\titlerunning{VLBA astrometry of Sharpless 269}
\author{L.~H.~Quiroga-Nu\~{n}ez 
\inst{1,2},
K.~Immer 
\inst{2},   
H.~J.~van~Langevelde 
\inst{2,1}, 
M.~J.~Reid 
\inst{3} 
\and 
R.~A.~Burns 
\inst{2} 
} 
\authorrunning{L.H. Quiroga-Nu\~{n}ez et al.}  
\institute{Leiden Observatory, Leiden University, P.O. Box 
9513, 2300 RA Leiden, The Netherlands.\\
\email{quiroganunez@strw.leidenuniv.nl} 
\and 
Joint Institute for VLBI ERIC (JIVE), 
Oude Hoogeveensedijk 4, 7991 PD, Dwingeloo, 
The Netherlands.
  \and 
Harvard-Smithsonian Center for Astrophysics, 60 Garden 
Street, Cambridge, MA 02138, USA.} 
\date{Received XX XXXX XXXX / Accepted XX XXXX XXXX}
\abstract
{Sharpless 269 ($\rm{S\,269}$) is one of a few HII regions in the outer spiral arm of the Milky Way with strong water maser emission. Based on data from the Very Long Baseline Interferometry (VLBI) Exploration of Radio Astrometry (VERA) array, two parallax measurements have been published, which differ by nearly $2\sigma$. Each distance estimate supports a different structure for the outer arm. Moreover, given its large Galactocentric radii, $\rm{S\,269}$ has special relevance as its proper motion and parallax have been used to constrain the Galactic rotation curve at large radii.}
{Using recent Very Long Baseline Array (VLBA) observations, we accurately measure the parallax and proper motion of the water masers in $\rm{S\,269}$. We interpret the position and motion of $\rm{S\,269}$ in the context of Galactic structure, and possible optical counterparts.}
{$\rm{S\,269}$'s 22 GHz water masers and two close by quasars were observed at 16 epochs between 2015 and 2016 using the VLBA. We calibrated the data by inverse phase referencing using the strongest maser spot. The parallax and proper motion were fitted using the standard protocols of the Bar and Spiral Structure Legacy survey.}
{We measure an annual parallax for $\rm{S\,269}$ of 0.241 $\pm$ 0.012 mas corresponding to a distance from the Sun of $4.15^{+0.22}_{-0.20}$ kpc by fitting four maser spots. The mean proper motion for $\rm{S\,269}$ was estimated as $0.16\pm0.26$ mas $\rm{yr^{-1}}$ and $-0.51\pm0.26$ mas $\rm{yr^{-1}}$ for $\mu_{\alpha} \ cos \delta$ and $\mu_{\delta}$ respectively, which corresponds to the motion expected for a flat Galactic rotation curve at large radius. This distance estimate, Galactic kinematic simulations and observations of other massive young stars in the outer region support the existence of a kink in the outer arm at $l \approx 140\degr$. Additionally, we find more than 2,000 optical sources in the Gaia DR2 catalog within 125 pc radius around the 3D position of the water maser emission; from those only three sources are likely members of the same stellar association that contains the young massive star responsible for the maser emission ($\rm{S\,269}$~IRS~2w).}
{}
\authorrunning{L.H. Quiroga-Nu\~{n}ez et al.} 
\titlerunning{VLBA astrometry of the Star-Forming Region 
$\rm{S\,269}$} 
\maketitle
\section{Introduction}
The Very Long Baseline Interferometry (VLBI) Exploration of Radio Astrometry (VERA\footnote{VERA is part of the National Astronomical Observatory of Japan}) project and the Bar and Spiral Structure Legacy (BeSSeL\footnote{\url{http://bessel.vlbi-astrometry.org/}}) survey have elucidated important aspects of the Milky Way galaxy, including values of its fundamental parameters and the nature of its spiral structure~\citep[][]{2011AN....332..461B,2014ApJ...783..130R,Honma2015,Sakai2015,Xu+Sci16}. The BeSSeL survey continues with additional VLBI observations of masers associated with High Mass Star-Forming Regions (HMSFRs) to better constrain the size and morphology of the Milky Way~\citep[see, e.g.,][]{Quiroga-Nunez2017,Sanna+Sci17}. This is relevant at large Galactocentric radii ($>12$ kpc), where only a few HMSFRs have been observed and their astrometric parameters are harder to measure~\citep[][and references within]{Hachisuka+apj17}. Also, the outer Galactic region is particularly interesting as it gauges the Galactic rotation curve, which is a crucial key to understand the role of dark matter in Galactic dynamics~\citep[see, e.g.,][]{Kent1986,Sofue2017}. 

In 2004, the VERA project started to monitor several maser bearing stars and star-forming regions to accurately determine their astrometric parameters~\citep{Honma13}. Their first result was the parallax and proper motion of the star-forming region Sharpless 269 ($\rm{S\,269}$), also known as Sh2-269, LBN 196.49$-$0.160 or G196.45$-$01.67~\citep{Honma2007a}. $\rm{S\,269}$ is a compact HII region in the outer Galaxy toward the Galactic anticenter at $l=196\fdg5$ and $b=-1\fdg7$~\citep{Sharpless1959}. It hosts several bright near-infrared (NIR) sources, in particular $\rm{S\,269}$~IRS~2w. This is a massive young O star with associated Herbig-Haro objects~\citep{Eiroa1994} and several species of masers~\citep{Minier2000,Sawada-Satoh2013a}. Water (22 GHz), methanol (6.7 and 12.2 GHz) and OH (1.6 GHz) maser emission around $\rm{S\,269}$~IRS~2w have been detected and studied for decades~\citep{1993LNP...412..279C,Minier2000,Lekht2001a} as the region presents signposts of star-forming activity~\citep{Jiang2003,Sawada-Satoh2013a} and intermediate scale interstellar turbulence~\citep{Lekht2001b}. $\rm{S\,269}$, therefore, represents one of a few well observed HII regions at large Galactocentric radii~\citep[$>13$ kpc,][]{Honma2007a}.

Using the VERA array,~\cite{Honma2007a} monitored the water maser emission from $\rm{S\,269}$~IRS~2w from 2004 to 2006. They reported strong maser emission of 480 Jy at 22 GHz with $V_\mathrm{LSR}\rm{= 19.7 \ km \ s^{-1}}$, and measured an annual parallax of 0.189 $\pm$ 0.008 mas, corresponding to a distance from the Sun of $5.28_{-0.22}^{+0.24}$ kpc and a Galactic rotational velocity similar to the Sun. This result suggested that the rotation curve of the Galaxy remains flat out to 13.5 kpc from the Galactic center~\cite[adopting $\rm{R_{\odot}}$ = 8.34 kpc,][]{2014ApJ...783..130R}.

Later \cite{Miyoshi2012b} and~\cite{Asaki2014a} disputed the distance to $\rm{S\,269}$ reported by~\cite{Honma2007a} , firstly pointing out that kinematic and optical photometric distance estimates reported shorter values~\citep[3.7-3.8 kpc, see][]{Moffat1979,WouterlootJ.G.A.;Brand1989,Xu2009}. Moreover, they reanalyzed the VERA data specifically using more compact maser spots than those used by~\cite{Honma2007a}, and reported a parallax value $0.247 \pm 0.034$ mas, which corresponds to a distance of $4.05_{-0.49}^{+0.65}$ kpc~\citep{Asaki2014a}. The tension between the two parallax distances is crucial for two reasons. First, the $\rm{S\,269}$ astrometric parameters have been used to constrain the Galactic rotation curve at large Galactocentric radii due to the limited number of sources with accurately measured distances in this area of the Galaxy. Second, the two distance estimates support a different structure of the outer spiral arm. The nearer distance estimate of $4.05_{-0.49}^{+0.65}$ kpc by~\cite{Asaki2014a} is inconsistent with previous distance estimates of the outer arm~\citep{Hachisuka+apj17}, suggesting a kink or bifurcation, whereas the larger distance of $5.28_{-0.22}^{+0.24}$ kpc by~\cite{Honma2007a} supports a smoother arm.

We now present the results and implications of a large number of recent Very Long Baseline Array (VLBA) observations of the $\rm{S\,269}$ region at 22 GHz. In Sect.~\ref{s_observations}, we describe the observations, the data reduction procedure and the methods used. The astrometric results and the search for optical members within the {\it Gaia} catalog are described in Sect.~\ref{s_results}. Then, in Sect.~\ref{s_discussion}, we analyze the maser emission structure, and the implications of the parallax and proper motion obtained regarding the structure of the outer arm and optical associations. Finally, we present the main conclusions of this work in Sect.~\ref{s_conclusions}. 
\section{Observations}
\label{s_observations}
From August 2015 to October 2016, we conducted 16 epochs of phase reference observations of water masers present in $\rm{S\,269}$, using two extragalactic continuum sources (J0613+1306 and J0619+1454) as close by position references at 0.73$^\circ$ and 1.67$^\circ$, respectively, from $\rm{S\,269}$. The observations were made using the VLBA operated by the National Radio Astronomy Observatory (NRAO\footnote{The National Radio Astronomy Observatory is a facility of the National Science Foundation operated under cooperative agreement by Associated Universities, Inc.}) under program BR210E. Table~\ref{t_epochstime} shows the dates and times of the 16 observations, which correspond to a sequence of four observations (i.e., one in late summer, two in late winter or early spring, and one more in the next late summer) repeated four times, close in time during each sequence.

Four adjacent 16 MHz bands, each in right and left circular polarization, were used with the third band centered on an $V_\mathrm{LSR}$ of 15 km $\rm{s^{-1}}$, assuming a rest frequency of the water maser $J_{K_{a}K_{c}}=6_{16} \rightarrow 5_{23}$ transition of 22,235.080 MHz. The observations were processed with the VLBI software correlator VLBA-DiFX\footnote{DiFX is developed as part of the Australian Major National Research Facilities Programme by the Swinburne University of Technology and operated under license.}, producing 2,000 and 32 spectral channels per band for the line and continuum data, respectively. This yielded a velocity spacing of 0.108 km $\rm{s^{-1}}$ for the line data. In addition, to estimate and then remove residual tropospheric delays relative to the correlator model, we inserted four geodetic blocks during each observational epoch~\citep[details about geodetic observations can be found in][]{Reid+Apj09a}. The observation cycles were designed such that $\rm{S\,269}$ was observed for 30 seconds (typically a 10 second slew and 20 seconds on source) followed by a compact extragalactic source for 30 seconds. Therefore, the center of the maser scans used for phase referencing were 60 seconds apart, which is shorter than the coherence time for VLBA observations at this frequency. Positions and flux densities for the dominant maser spot ($V_\mathrm{LSR}=19.6 \ \rm{km \ s^{-1}}$), used as the phase reference for the extragalactic sources, and the two continuum sources are shown in Table~\ref{t_sourcesinfo}. The data reduction was performed using the NRAO Astronomical Image Processing System (AIPS), together with scripts written in ParselTongue~\citep{Kettenis2006}, following standard BeSSeL survey data reduction methods~\cite[see][]{Reid+Apj09a}.

\begin{table}
\caption{Information of the strongest $\rm{S\,269}$ maser spot detected, and both extragalactic sources used for parallax and proper motion estimate.\label{t_sourcesinfo}}
\begin{center}
\begin{tabular}{cccc}
\hline \hline
\multicolumn{1}{c}{Source} & $\alpha$ (J2000) & $\delta$ (J2000) &$S_{\nu}$ (Jy \\
Name &(hh:mm:ss)&  ($^\circ$ : $^\prime$ : $^{\prime\prime}$) & $\rm{beam^{-1}}$) \\
\hline
$\rm{S\,269}$&06:14:37.6410&+13:49:36.6930&95.8 \\
J0613+1306&06:13:57.6928&+13:06:45.4010&0.2 \\
J0619+1454&06:19:52.8723&+14:54:02.7346&0.1 \\
\hline
\end{tabular} \\
\end{center}
{\footnotesize {\bf Notes.} $\rm{S\,269}$ spot is shown in Fig.~\ref{p_elongated} and corresponds to spot \RNum{1} in Fig.~\ref{p_featuresdist_asa_honma}. The peak flux density corresponds to the observations made at Epoch H.}
\end{table}

\begin{figure*}
\centering
\includegraphics[width=18.0cm]{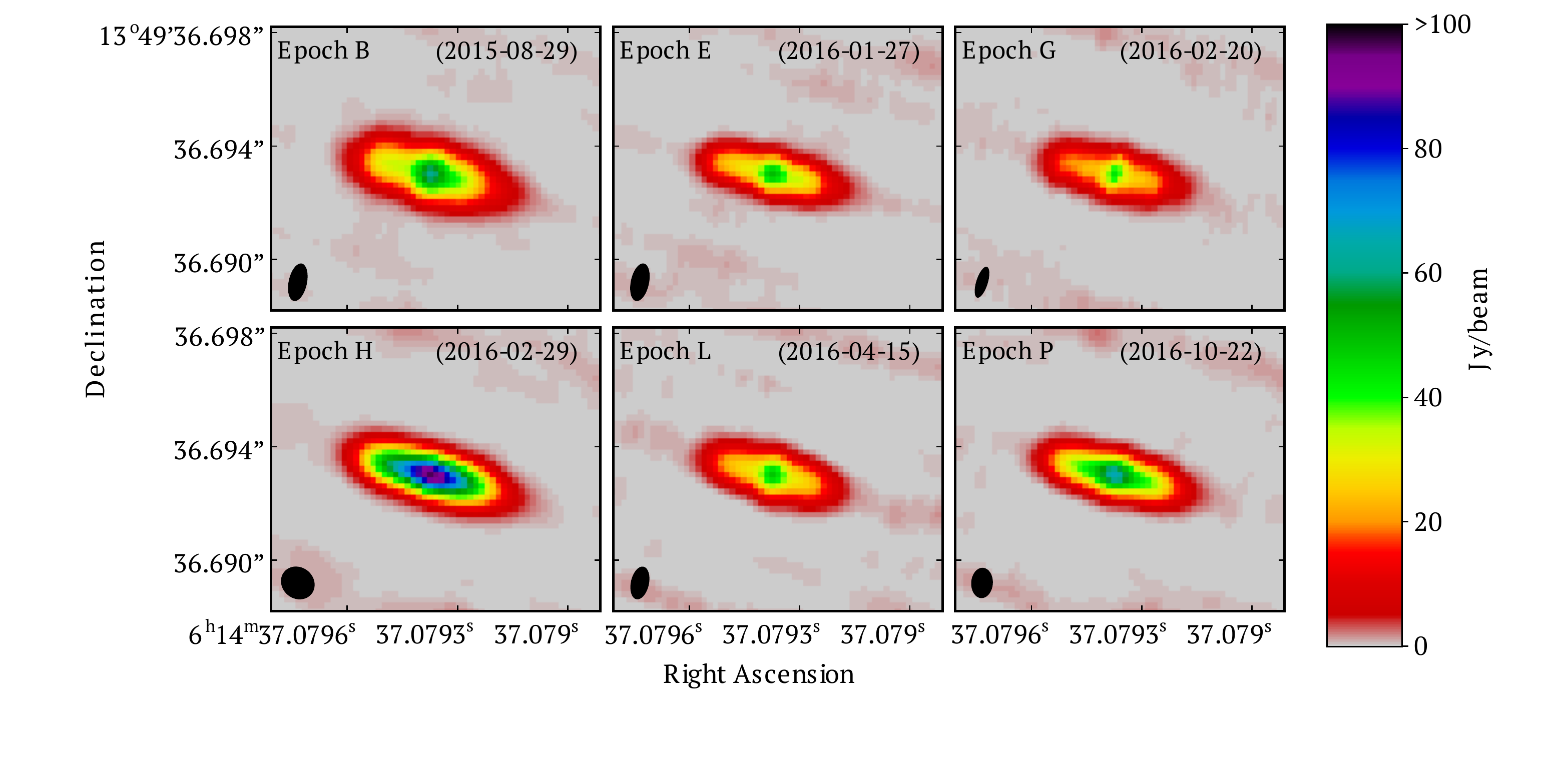}
\caption{Strongest 22 GHz maser spot in $\rm{S\,269}$ region showing an elongated shape at six different epochs in the VLBA observations between 2015 and 2016. Based on its position and $V_\mathrm{LSR}$ (but not its motion, see Sect.~\ref{s_discussion_elongated}), it seems to correspond to the strongest maser spot reported by~\cite{Honma2007a}, using VERA observations in 2004 and 2005. This maser spot corresponds to the spot \RNum{1} in Fig.~\ref{p_featuresdist_asa_honma}. The shape and size of the beam for each epoch is shown in the bottom left corner.}
\label{p_elongated}
\end{figure*}

\begin{figure}
\centering
\includegraphics[scale=0.41]{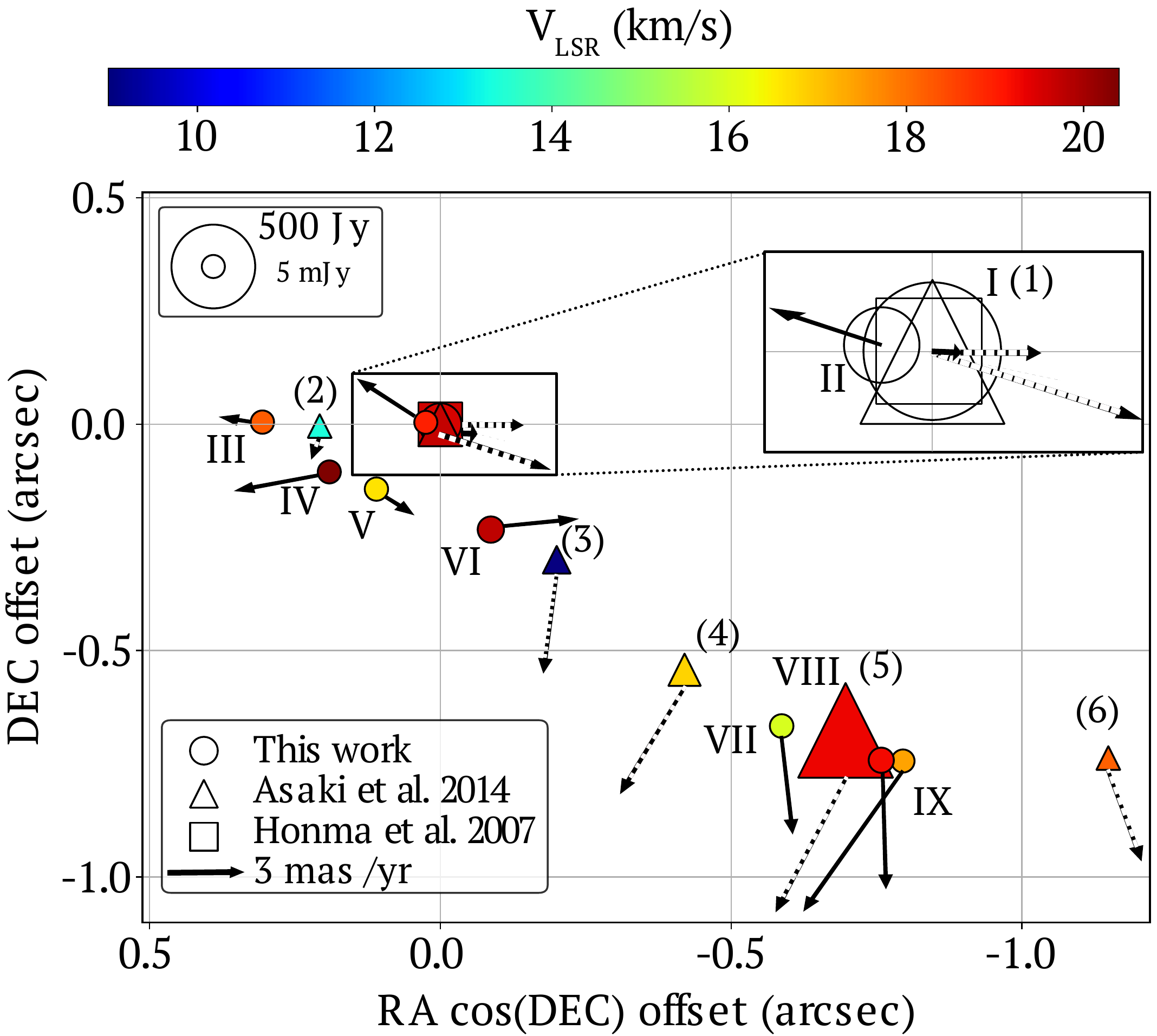}
\caption{  Distribution of water maser spots around the strongest maser emission (where several maser spots coincide, see zoom in) with their radial velocity values expressed using a color scheme indicated by the bar at the top. The maser spots detected by the VLBA in 2015-2016 (see Table~\ref{t_feautres}) are shown as circles, with Roman numbers and their proper motion as continuum arrows. Whereas, the main maser spots detected by VERA in 2004-2005 and published by~\cite{Honma2007a} and~\cite{Asaki2014a} are shown as squares and triangles, with Arabic numerals and their proper motion as dashed arrows. The size of the markers is proportional to the flux density peak of each maser spot (see upper left corner convention).} 
\label{p_featuresdist_asa_honma}%
\end{figure}

\begin{table*}[tbp]
\caption{Relative position, radial velocities and flux density peaks at certain observational epoch for the water maser spots shown in Fig.~\ref{p_featuresdist_asa_honma}.\label{t_feautres}}
\begin{center}
\begin{tabular}{rrrrrr}
\hline \hline
\multicolumn{1}{c}{Spot ID}& \multicolumn{1}{c}{$\Delta \alpha$}& \multicolumn{1}{c}{$\Delta \delta$}& \multicolumn{1}{c}{$V_\mathrm{LSR}$}& \multicolumn{1}{c}{$F_{max}$}& \multicolumn{1}{c}{Peak}   \\
\multicolumn{1}{c}{(Fig.~\ref{p_featuresdist_asa_honma})} & \multicolumn{1}{c}{(mas)}&\multicolumn{1}{c}{(mas)}&\multicolumn{1}{c}{$\rm{(km \ s^{-1})}$}&\multicolumn{1}{c}{$\rm{(Jy \  beam^{-1})}$}&\multicolumn{1}{c}{epoch}\\
\hline
\RNum{1} &  $-$0.101 $\pm$0.005& $-$0.020$\pm$0.002&19.6&95.8&H\\
\RNum{2} & 24.984 $\pm$0.044&4.116$\pm$0.024&19.0&0.5&H\\
\RNum{3} & 305.924$\pm$0.007&4.652$\pm$0.012&18.2& 1.4&B\\
\RNum{4} & 190.894$\pm$0.025& $-$105.335$\pm$0.044&20.4& 0.3&D\\
\RNum{5} & 109.886$\pm$0.006& $-$143.253$\pm$0.010&16.6& 1.4&O\\
\RNum{6} &  $-$86.811$\pm$0.080& $-$232.680$\pm$0.062&19.8&12.9&H\\
\RNum{7} & $-$586.729$\pm$0.011& $-$666.299$\pm$0.005&16.0& 2.5&P\\
\RNum{8} & $-$758.007$\pm$0.009& $-$741.869$\pm$0.013&19.2&8.5&O\\
\RNum{9} & $-$795.345$\pm$0.015& $-$743.615$\pm$0.022&17.4& 0.6&A\\
\hline
\end{tabular}
\end{center}
{\footnotesize
{\bf Notes.} The offsets were measured with respect to the strongest water maser spot, for which the absolute position is given in Table~\ref{t_sourcesinfo}. The epoch of fit was taken as the middle time of the VLBA observations, which is 2016.2.}
\end{table*}

\section{Results}
\label{s_results}

Sixteen data cubes were constructed (one per epoch), each measuring 32,768 pixels $\times$ 32,768 pixels $\times$ 300 channels. This corresponds to an image of $1.64^{\prime\prime} \times 1.64^{\prime\prime}$ using a cellsize of 0.05 mas $\rm{pixel^{-1}}$ within a radial velocity range between $-6.4$ and 25.7 $\rm{km \ s^{-1}}$. The range values for the data cube were calculated to include all the maser spots reported in~\cite{Miyoshi2012b} and~\cite{Asaki2014a}.

\begin{figure*}
\centering
\includegraphics[width=16.0cm]{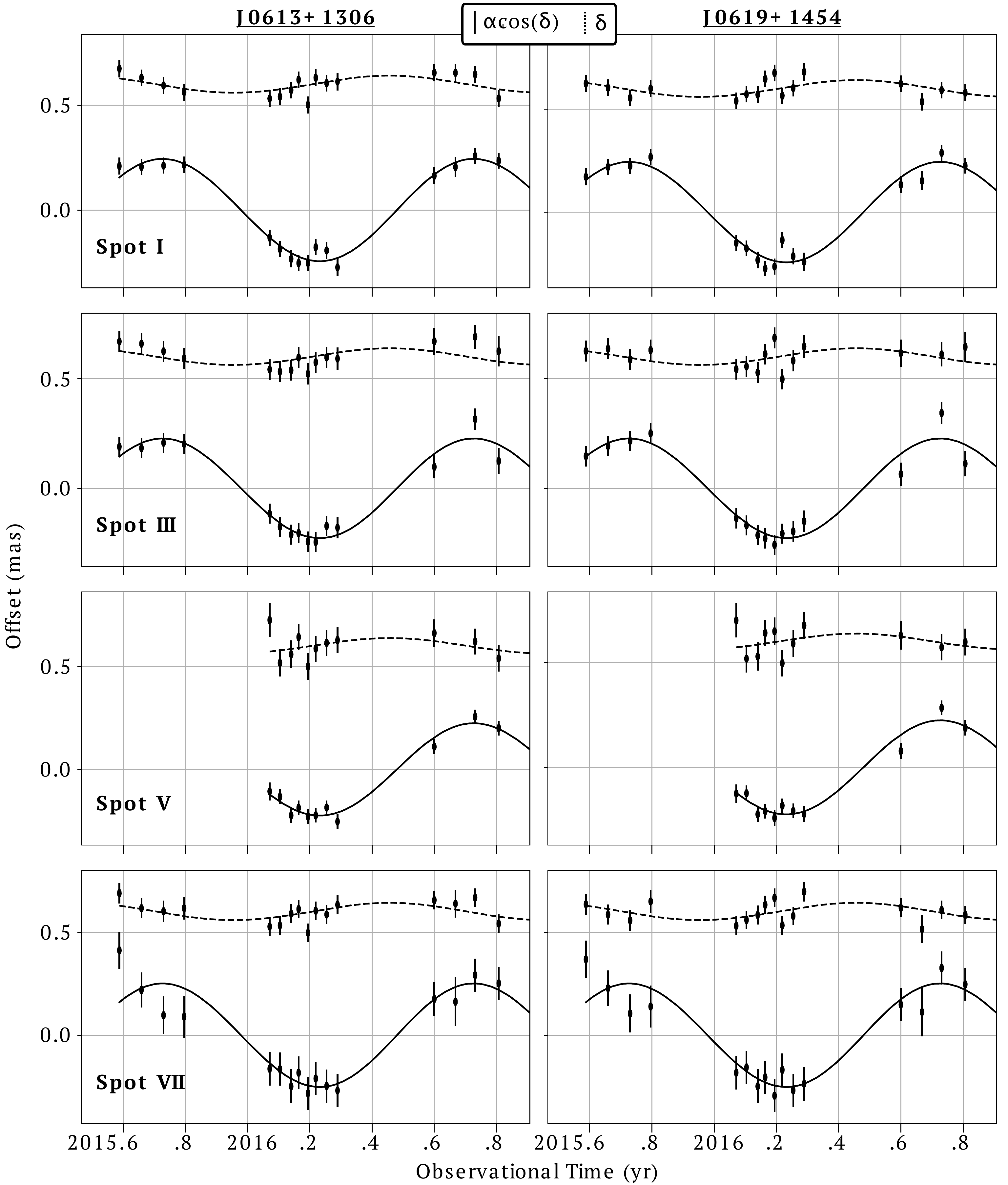}
\caption{Astrometric offsets for four different water masers with respect to the quasars used as reference position: J0613+1306 in the left plots and J0619+1454 in the right plots. The proper motions were subtracted from the parallax signatures. The solid and dashed lines represent the eastward ($\rm{\alpha \ cos\delta}$) and northward ($\rm{\delta}$) individual fitting listed in Table~\ref{t_spot_pi_ppm_fit}, respectively. The Northward offset ($\rm{\delta}$) was shifted +0.6 mas for visualization purposes.}
\label{p_offset_pifit_individual}%
\end{figure*}

\begin{table*}[ht]
\caption{Fitting results of parallax and proper motion for four 22 GHz water maser spots present in $\rm{S\,269}$ with respect to two extragalactic continuum sources, under the usual assumption that the quasars are stationary during our observations. Radial velocities of those masers are also shown. The description of how the combined parallax and proper motion were obtained is described in Sect.~\ref{s_results_astrometric}.}
\label{t_spot_pi_ppm_fit}
\resizebox{\hsize}{!}
{
\begin{tabular}{cccrrcrr}
\hline \hline
Maser & Quasar & $\pi$& \multicolumn{1}{c}{$\rm{\mu_{{\alpha}}\ cos \delta}$}& \multicolumn{1}{c}{$\rm{\mu_{{\delta}}}$} & $V_\mathrm{LSR}$ & \multicolumn{1}{c}{Combined} & \multicolumn{1}{c}{Combined} \\
Spot &  & (mas) & \multicolumn{1}{c}{(mas yr$^{-1}$)}  & \multicolumn{1}{c}{(mas yr$^{-1}$)} & (km $\rm{s}^{-1}$) & \multicolumn{1}{c}{$\rm{\mu_{{\alpha}}\ cos \delta}$ (mas yr$^{-1}$)} & \multicolumn{1}{c}{ $\rm{\mu_{{\delta}}}$ (mas yr$^{-1}$)} \\
\hline
\multirow{2}{*}{\RNum{1}} & J0613+1306& 0.244$\pm$ 0.010  &  $-$0.091 $\pm$ 0.024  &  $-$0.003 $\pm$ 0.029 & \multirow{2}{*}{19.6} & \multirow{2}{*}{ $-0.099 \pm 0.019$} & \multirow{2}{*}{ $-0.008 \pm 0.020$} \\
& J0619+1454&  0.243 $\pm$ 0.013 &  $-$0.107 $\pm$ 0.032&  $-$0.012 $\pm$ 0.028 & & &\\
\hline
\multirow{2}{*}{\RNum{2}} & J0613+1306& \multirow{2}{*}{$-$}  &  0.518 $\pm$ 0.187  &  0.408 $\pm$ 0.099 & \multirow{2}{*}{19.0} & \multirow{2}{*}{$0.507 \pm 0.128$} & \multirow{2}{*}{ $0.438 \pm 0.076$} \\
& J0619+1454&  &  0.497 $\pm$ 0.187&  0.464 $\pm$ 0.124 & & &\\
\hline
\multirow{2}{*}{\RNum{3}}& J0613+1306&0.226 $\pm$ 0.013 &  0.173 $\pm$ 0.033  &  0.049 $\pm$ 0.039 &  \multirow{2}{*}{18.2} & \multirow{2}{*}{$0.170 \pm 0.025$} & \multirow{2}{*}{$0.063 \pm 0.028$} \\
& J0619+1454& 0.228 $\pm$ 0.017 & 0.166 $\pm$ 0.041&0.078 $\pm$ 0.042 & & &\\
\hline
\multirow{2}{*}{\RNum{4}} & J0613+1306& \multirow{2}{*}{$-$}  &  0.767 $\pm$ 0.115  &  $-$0.234 $\pm$ 0.194 & \multirow{2}{*}{20.4} & \multirow{2}{*}{$0.763 \pm 0.089$} & \multirow{2}{*}{ $-0.185 \pm 0.139$} \\
& J0619+1454&  &  0.762 $\pm$ 0.144&  $-$0.137 $\pm$ 0.206 & & &\\
\hline
\multirow{2}{*}{\RNum{5}} & J0613+1306& 0.234  $\pm$ 0.033 &  $-$0.189 $\pm$ 0.032  &  $-$0.177 $\pm$  0.057 & \multirow{2}{*}{16.6} & \multirow{2}{*}{ $-0.187 \pm 0.052$} & \multirow{2}{*}{ $-0.177 \pm 0.057$} \\
& J0619+1454& 0.214  $\pm$ 0.042 &  $-$0.108 $\pm$ 0.133  &  $-$0.207 $\pm$ 0.085 & & &\\
\hline
\multirow{2}{*}{\RNum{6}} & J0613+1306& \multirow{2}{*}{$-$}  &  $-$0.639 $\pm$ 0.215  &  0.085 $\pm$ 0.068 & \multirow{2}{*}{19.8} & \multirow{2}{*}{$-0.650 \pm 0.147$} & \multirow{2}{*}{ $0.080 \pm 0.050$} \\
& J0619+1454&  &  -0.661 $\pm$ 0.212&  0.069 $\pm$ 0.082 & & &\\
\hline
\multirow{2}{*}{\RNum{7}} & J0613+1306  
& 0.254 $\pm$ 0.026  &  $-$0.108 $\pm$ 0.064& $-$1.255 $\pm$ 0.034 &
\multirow{2}{*}{16.0} & \multirow{2}{*}{ $-0.111 \pm 0.043$} & \multirow{2}{*}{ $-1.254 \pm 0.024$} \\
& J0619+1454& 0.248 $\pm$ 0.024&  $-$0.115 $\pm$ 0.060& $-$1.254 $\pm$ 0.035 & & & \\   
\hline
\multirow{2}{*}{\RNum{8}} & J0613+1306& \multirow{2}{*}{$-$}  &  $-$0.209 $\pm$ 0.286  &  $-$1.602 $\pm$ 0.490 & \multirow{2}{*}{19.2} & \multirow{2}{*}{$-0.031 \pm 0.160$} & \multirow{2}{*}{ $-1.580 \pm 0.250$} \\
& J0619+1454&  &  0.152 $\pm$ 0.120&  $-$1.542 $\pm$ 0.347 & & &\\
\hline
\multirow{2}{*}{\RNum{9}} & J0613+1306& \multirow{2}{*}{$-$}  &  0.950 $\pm$ 0.819  &  $-$2.142 $\pm$ 0.213 & \multirow{2}{*}{17.4} & \multirow{2}{*}{$1.052 \pm 0.529$} & \multirow{2}{*}{ $-1.959 \pm 0.139$} \\
& J0619+1454&  &  1.149 $\pm$ 0.839&  $-$1.772 $\pm$ 0.140 & & &\\
\hline
\hline
Combined & \multirow{2}{*}{} & \multirow{2}{*}{0.241 $\pm$ 0.012} &\multirow{2}{*}{} & Average&\multirow{2}{*}{}&\multirow{2}{*}{0.157 $\pm$ 0.066}&\multirow{2}{*}{-0.509 $\pm$ 0.037} \\
Parallax& & & &Proper Motion& & & \\
\hline
\end{tabular}
}
{\footnotesize
{\bf Notes.} The first column gives the maser spot number used in Fig.~\ref{p_featuresdist_asa_honma}. The annual parallax values ($\pi$) provided in the third column were fitted for only four masers (see Sect.~\ref{s_results_astrometric}). For the absolute proper motion fittings (columns 4,5,7,8), we fixed the parallax value in 0.241 $\pm$ 0.012 mas. The procedures to obtain the combined and average values reported in this table are described in Sect~\ref{s_results_astrometric}.}
\end{table*}

We detected nine maser spots that were persistent for at least three epochs. Gaussian brightness distributions were fitted to the maser images by a least-squares method using the task $JMFIT$ within AIPS. Table~\ref{t_feautres} shows the results of the fitting for each maser spot, together with its radial velocity and the maximum flux density across all epochs. Figure~\ref{p_elongated} shows the strongest water maser detected at representative epochs, while Fig.~\ref{p_featuresdist_asa_honma} shows the distribution of the maser spots, proper motion and radial velocities found in our VLBA data together with those reported in~\cite{Honma2007a} and~\cite{Asaki2014a}. The strongest maser spot is labeled as~``\RNum{1}'' and it was used as central reference. In the VLBA observations, the water masers are confined to a radial velocity range between 16.0 and 20.4 $\rm{km \ s^{-1}}$. This is within the velocity range found in single dish spectra for S269~\citep{Lekht2001a}.
\subsection{Elongated water maser emission}

The strongest maser spot was detected in all sixteen epochs with a flux density maximum of 95.8 $\rm{Jy \ beam^{-1}}$ at epoch H. This spot has a distinctive elongated shape at all epochs (see Fig.~\ref{p_elongated}), as was firstly reported by~\cite{Miyoshi2012b} and highlighted by~\cite{Asaki2014a}. The spot varies somewhat over time but retains its basic shape throughout all our observations. 

The inner core for this elongated spot could be well fitted by a single, compact, Gaussian brightness distribution, and we used the AIPS task $JMFIT$ with a 2 mas box to fit the core.
\subsection{Astrometric measurements for $\rm{S\,269}$}
\label{s_results_astrometric}
Only four of the nine 22 GHz water maser spots were detected in at least ten epochs, which allow a robust fitting for the annual parallax sinusoidal signature in right ascension and declination. We also added error floor values to the position uncertainties in both sky coordinates and adjusted them to obtain $\chi_\nu^2 \approx 1$~\citep[see details in][]{Reid+Apj09a}.

The four maser spots used in the parallax fitting are labeled in Fig.~\ref{p_featuresdist_asa_honma} as spots~\RNum{1},~\RNum{3},~\RNum{5} and~\RNum{7} and were detected in 16, 15, 10 and 16 epochs, respectively. As the four spots gave consistent parallax results (including the elongated spot, see Table~\ref{t_spot_pi_ppm_fit} and Figure~\ref{p_offset_pifit_individual}), we also have calculated a combined fit by simultaneously fitting all data (i.e., four spots measured for both quasars). This yields a combined annual parallax value of $0.241 \pm 0.012$ mas. The uncertainty in the parallax includes an additional scaling factor of $\sqrt{N}$, where $N$ is the number of maser spots used for the fit. This accounts for the correlated systematic position variations among maser spots caused by atmospheric effects~\citep{Reid+Apj09a}.

To estimate an average proper motion of the region, we fixed the annual parallax (previously calculated with only four maser spots) and fit the proper motions for the nine masers with respect to both (labeled as combined in Table~\ref{t_spot_pi_ppm_fit})  continuum extragalactic sources. Then, we averaged the proper motions of all nine maser spots detected by a standard mean (labeled as average proper motion in Table~\ref{t_spot_pi_ppm_fit}). Moreover, we include a 5 $\rm{km \ s^{-1}}$ uncertainty that accounts for the uncertainty of the motion of the masers with respect to the center of mass of the HMSFR. We note that the quiescent gas has $V_\mathrm{LSR}$ similar to the masers~\citep[i.e., 17.7, 16.5 and 18.2 $\rm{km \ s^{-1}}$ for CO, $\lbrack$SII$\rbrack$ and HCN, respectively,][]{Carpenter1990,Godbout1997,Pirogov1999}. Finally, we estimate $\rm{\mu_{{\alpha}}\ cos \delta=0.16\pm 0.26 \ mas \ yr^{-1}}$ and $\rm{\mu_{{\delta}}=-0.51 \pm 0.26 \ mas \ yr^{-1}}$ for the average proper motion of $S\,269$.

\begin{figure*}
\centering
\resizebox{\hsize}{!}{\includegraphics{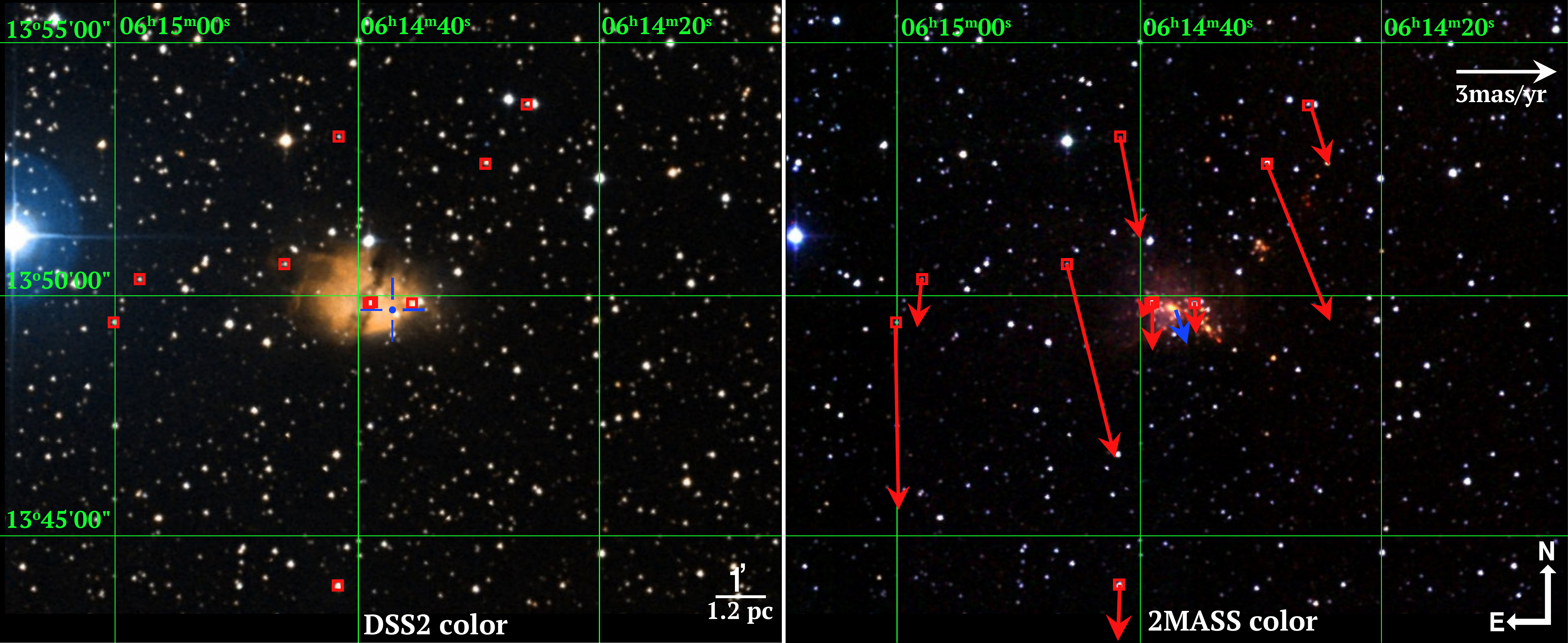}}
\caption{Image of the $\rm{S\,269}$ region using data from DSS2 {\bf (left panel)} and 2MASS {\bf (right panel)} with a sky projected size of $15.65^\prime \times 12.33^\prime$, centered on the water maser emission (blue central sign) detected with the VLBA (Table~\ref{t_sourcesinfo}), that seems to be triggered by the massive young star $\rm{S\,269}$~IRS~2w~\citep{Asaki2014a}. In the left panel, the molecular cloud appears as a butterfly with two wings separated by a dark fringe where a B0.5 star has been detected in the center~\citep{Moffat1979}. In contrast, near-infrared observations (right panel) show that the center of the clustered region of massive young objects coincides with the position of $\rm{S\,269}$~IRS~2w. The red squares are the closest optical sources around $\rm{S\,269}$ found within the {\it Gaia} DR2 catalog (see Sects.~\ref{s_Gaiaresult} and~\ref{s_optical_discussion_hrd}) and their astrometric information is shown in Table~\ref{t_gaia_sources}. The red arrows display the proper motions for such sources, whereas the blue arrow is the proper motion calculated by us using the water masers in the region. The images were generated with the Aladdin interface~\citep{Bonnarel2000}, where the color map descriptions of DSS2 and 2MASS can be found through~\url{alasky.u-strasbg.fr/DSS/DSSColor/} and~ \url{alasky.u-strasbg.fr/2MASS/Color/}.}
\label{p_Gaiacounter_S269_dss}%
\end{figure*}

\subsection{Cross-matching with {\it Gaia} DR2}
\label{s_Gaiaresult}
$\rm{S\,269}$~IRS~2w is located in a CO molecular cloud with a projected size of $7^\prime \times 10^\prime$~\citep{Heydari-Malayeri1982,Carpenter1990}. Other massive young stars, which could belong to the same stellar association, are expected to be detected in the vicinity of the CO molecular cloud. Since molecular gas is mostly confined to the Galactic plane (and mainly in the Galactic spiral arms) in a layer with FWHM of several hundred pc for Galactic radii greater than 10 kpc~\citep{Heyer2015}, we searched in the {\it Gaia} DR2 catalog within a spherical region around $\rm{S\,269}$'s location in 3D. As GMCs usually extend from 5 pc up to 120 pc, with a very few exceptional cases over 150 pc~\citep{Murray2011}, we used a radius 125 pc ($1\fdg73$ at $\rm{S\,269}$'s distance) as a conservative value to guarantee that most of the plausible sources associated with the S269 region were included in the inspected range. This corresponds to a parallax range (including $\pm\sigma$) from 0.2225 to 0.2615 mas. Figure~\ref{p_Gaiacounter_S269_dss} shows the $\rm{S\,269}$ region using data from the Digital Sky Survey 2~\citep[DSS2\footnote{\url{http://archive.eso.org/dss/dss}}][]{Lasker94} and the Two Micron All Sky Survey~\citep[2MASS\footnote{\url{https://www.ipac.caltech.edu/2mass/}}][]{Skrutskie2006} centered on the maser emission. 

We only selected sources with confident parallax measurements in {\it Gaia} DR2 (i.e. $\sigma_{\pi}/\pi<0.2$) that allow direct distance estimates~\citep{Bailer-Jones2015}. In total, there are 2,279 sources that fall into the spherical region defined. The closest ten {\it Gaia} counterparts in 3D are highlighted in red in Fig.~\ref{p_Gaiacounter_S269_dss}, and their astrometric information is shown in Table~\ref{t_gaia_sources}. We did not find an optical counterpart in {\it Gaia} DR2 that corresponds to the massive young star which surrounding material is yielding the water maser emission detected at 22 GHz. This is expected for a newly forming star that is deeply embedded in its placental material. However, the three closest optical counter parts (first three rows in Table~\ref{t_gaia_sources}) were found within a core size of the $\rm{S\,269}$ HII region ($\rm{3.9 \ pc \times 2.8 \ pc}$) estimated by~\cite{Godbout1997}. These three {\it Gaia} DR2 sources have an average parallax that differs with respect to the VLBA observations by $-32\pm23$ $\mu$as (assuming a {\it Gaia} zero-point correction of $\sim$ $-0.03$ mas) and an average proper motion that differs by $0.02\pm0.65$ and $-0.16\pm0.77$ $\rm{mas \ yr^{-1}}$ for $\rm{\mu_{{\alpha}}\ cos \ \delta}$ and $\rm{\mu_{{\delta}}}$ respectively. Therefore, they are likely members of the stellar association that contains $\rm{S\,269}$~IRS~2w.
\section{Discussion}
\label{s_discussion}
\subsection{Long-lived and extended water maser emission}
\label{s_discussion_elongated}

\subsubsection{Elongated maser spot}

The unusual morphology of the spot~\RNum{1} over many (from VERA in 2004 to VLBA in 2016) observations  (Fig.~\ref{p_elongated}) compared to typical maser spots vouches for its authenticity. An instrumental artifact, instead, would manifest a similar structural appearance in all similarly calibrated maser emission in the data cube, which is not the present case. We further analyzed the maser structure using DIFmap's modelfit and projplot tools. The structure of the maser is well fit by an elongated structure (P.A.~$\sim$78$\degr$) plus a compact core as it is evident in Fig.~\ref{p_elongated}.

\subsubsection{Correspondence with previous water maser observations}

As a collisionally pumped maser transition, 22 GHz water masers are typically found in turbulent regions of post-shocked gas associated with star formation outflows~\citep[see, e.g.,][ and the references within]{Liljestro+2000,Hollenbach2013}. While a shocked region at some radial velocity may consistently produce maser emission around the shock velocity, the individual maser spots are typically short-lived~\citep[$\lesssim$1 yr, see e.g.,][]{Tarter1986}. With this in mind, the persistent appearance of the maser spot~\RNum{1} in Fig.~\ref{p_featuresdist_asa_honma} at about the same location in the source makes it remarkable. Its position and shape seem to correspond to the maser spot reported by~\cite{Honma2007a} for the VERA observations made between 2004 and 2005, and the reanalysis of VERA data made by~\cite{Miyoshi2012b} and~\cite{Asaki2014a}. Its longevity may be related to its complex structure, maser spots typically being much more compact. In principle, this could allow us to fit the parallax and proper motion over a 10 year baseline for this spot.

Although this region seems to persistently yield elongated maser emission over decades, indicating it is the same masering cloud, there is a significant difference between the proper motion measured for the observing set in 2004 using VERA and 2015 using the VLBA, with respect to the same quasar (J0613+1306). On the one hand, \cite{Honma2007a} reported for the elongated maser spot $\rm{\mu_{\alpha} \  cos \delta = -0.388 \pm 0.014 \ mas \ yr^{-1}}$ and $\rm{\mu_{\delta} = -0.118 \pm 0.071 \ mas \ yr^{-1}}$, and~\cite{Asaki2014a} reported $\rm{-0.738 \pm 0.008 \ mas \ yr^{-1}}$ and $\rm{-0.249 \pm 0.007 \ mas \ yr^{-1}}$~\citep[Spot ID 6 in][]{Asaki2014a}. In contrast, we estimated a much slower proper motion of $\rm{\mu_{\alpha} \ cos \delta = -0.099 \pm 0.019 \ mas \ yr^{-1}}$ and $\rm{\mu_{\delta} = -0.008 \pm 0.020 \ mas \ yr^{-1}}$. It seems likely that small changes in the coherent amplification path of the maser account for these differences. Thus, we cannot fit the parallax and proper motion for this spot on a 10 year baseline given its velocity discrepancy, but its peculiar morphology hints that those spots represent the same maser region. 

\subsubsection{Source of the elongated water maser emission}

In principle, an amplified background source could mimic the particular properties (morphology and longevity) of the elongated maser spot, however, there is no sign of such continuum source when we inspected the continuum bands of our observations. Alternatively, the linear distribution of the water maser spots (Fig.~\ref{p_featuresdist_asa_honma}) suggests that we are observing the front shock of the outflow moving in the southeastern direction. This direction is confirmed by infrared data from the Simultaneous-3color InfraRed Imager for Unbiased Survey (SIRIUS), where $\rm{S\,269}$~IRS~2 shows a bipolar jet in the southeastern-northwestern direction~\citep[but it remains unclear if it is associated with $\rm{S\,269}$~IRS~2w or $\rm{S\,269}$~IRS~2e]{Jiang2003}. This fact suggests that material may have been compressed yielding an elongated maser emission, which indeed is perpendicular to the shock motion. Moreover, CO, [SII] and HCN observations of the $\rm{S\,269}$ region reported a $V_\mathrm{LSR}$ of 17.7, 16.5 and 18.2$~\rm{km \ s^{-1}}$, respectively~\citep{Carpenter1990,Godbout1997,Pirogov1999}, which differs with respect to our maser observations supporting the jet origin of the maser emission.

\subsubsection{Cyclic maser emission in $\rm{S\,269}$}

\cite{Lekht2001a} monitored the water maser emission toward $\rm{S\,269}$ for more than 20 years (1980-2001) using the 22 meter telescope of the Pushchino Radio Astronomy Observatory. They reported a $V_\mathrm{LSR}$ range of $\rm{[19.6-20.4]\ km \ s^{-1}}$ in which our VLBA observations and also those made by~\cite{Honma2007a} fall. Although this single dish effort could not image the water maser, they found that the integrated flux of the strongest maser had cyclic emission between 70 and 600 Jy with a period of between 4.8 and 6.6 years. Assuming the cyclic emission suggested by~\cite{Lekht2001a}, subsequent peak emissions (over 200 Jy) should have occurred between [2004.4-2006.2], [2009.2-2012.8] and [2014-2019.4]. Both VERA~\citep{Honma2007a} and VLBA (this work) observations spanned more than one year within these time ranges, but only VERA observations showed enhanced emission of 480 Jy. Although there is evidence of previous flares at radio wavelengths in $\rm{S\,269}$~\citep[see, e.g.,][]{1993LNP...412..279C}, the cyclic emission proposed by~\cite{Lekht2001a} does not seem consistent with our VLBA observations. 

\subsection{$\rm{S\,269}$ astrometric parameters}
\label{s_distance_pecuvel}

\begin{figure}
\centering
\resizebox{\hsize}{!}{\includegraphics{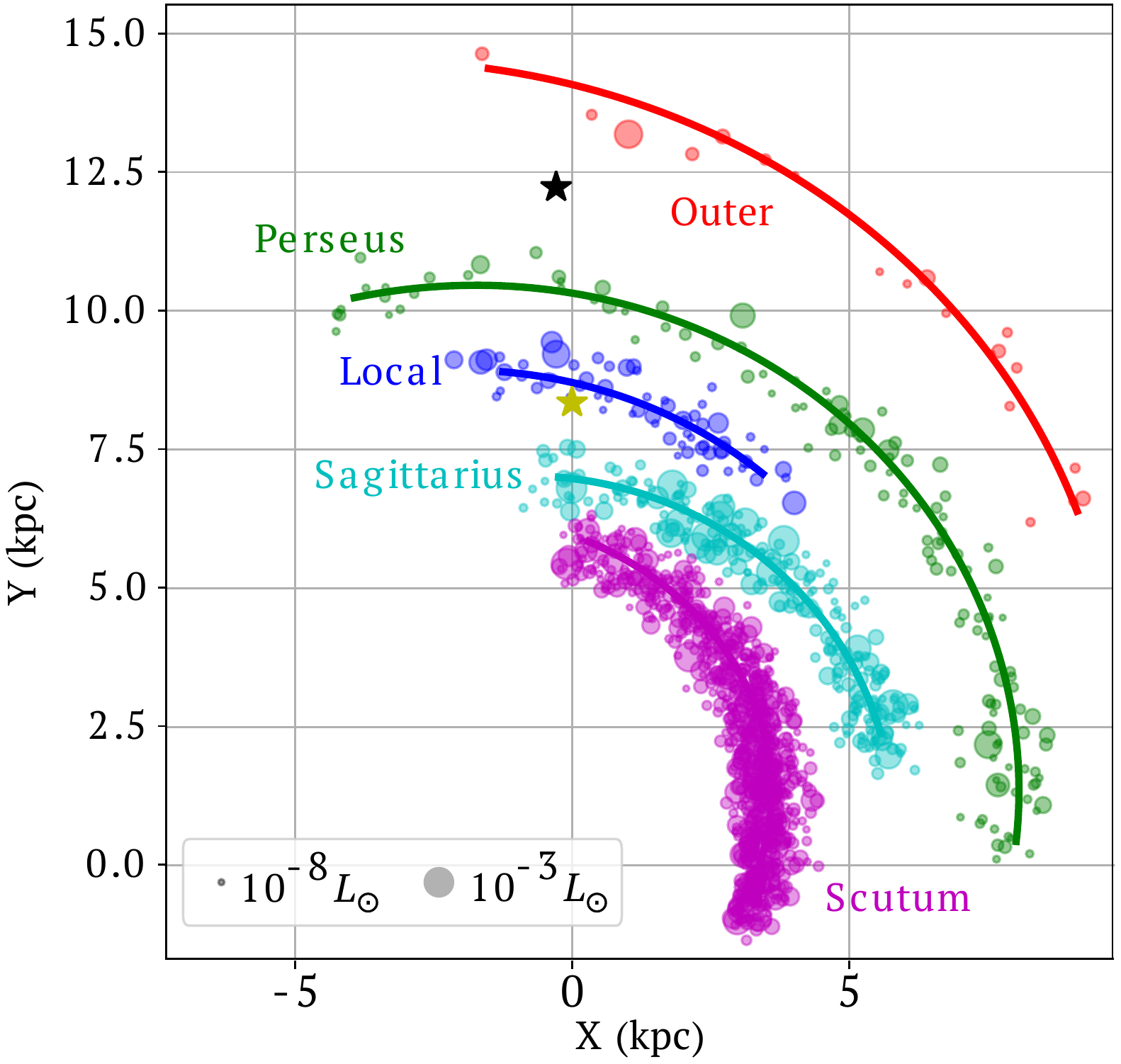}}
\caption{Plan view of a simulation of the Galactic maser distribution for maser bearing stars around the spiral structure using the model developed by~\cite{Quiroga-Nunez2017}. The spiral structure estimated by~\cite{2014ApJ...783..130R} was populated with artificial sources to compare the phase-space density distribution of the spiral arms with $\rm{S\,269}$ properties (see Sect.~\ref{s_membership_arm}). The Galactic center is located at (0,0), and the yellow and black stars correspond to the position of the Sun~\citep{2014ApJ...783..130R} and $\rm{S\,269}$, respectively, where their positional error bars are smaller than the size of the marker.}
\label{p_new_Arms}%
\end{figure}

\begin{figure*}
\centering
\includegraphics[width=16.0cm]{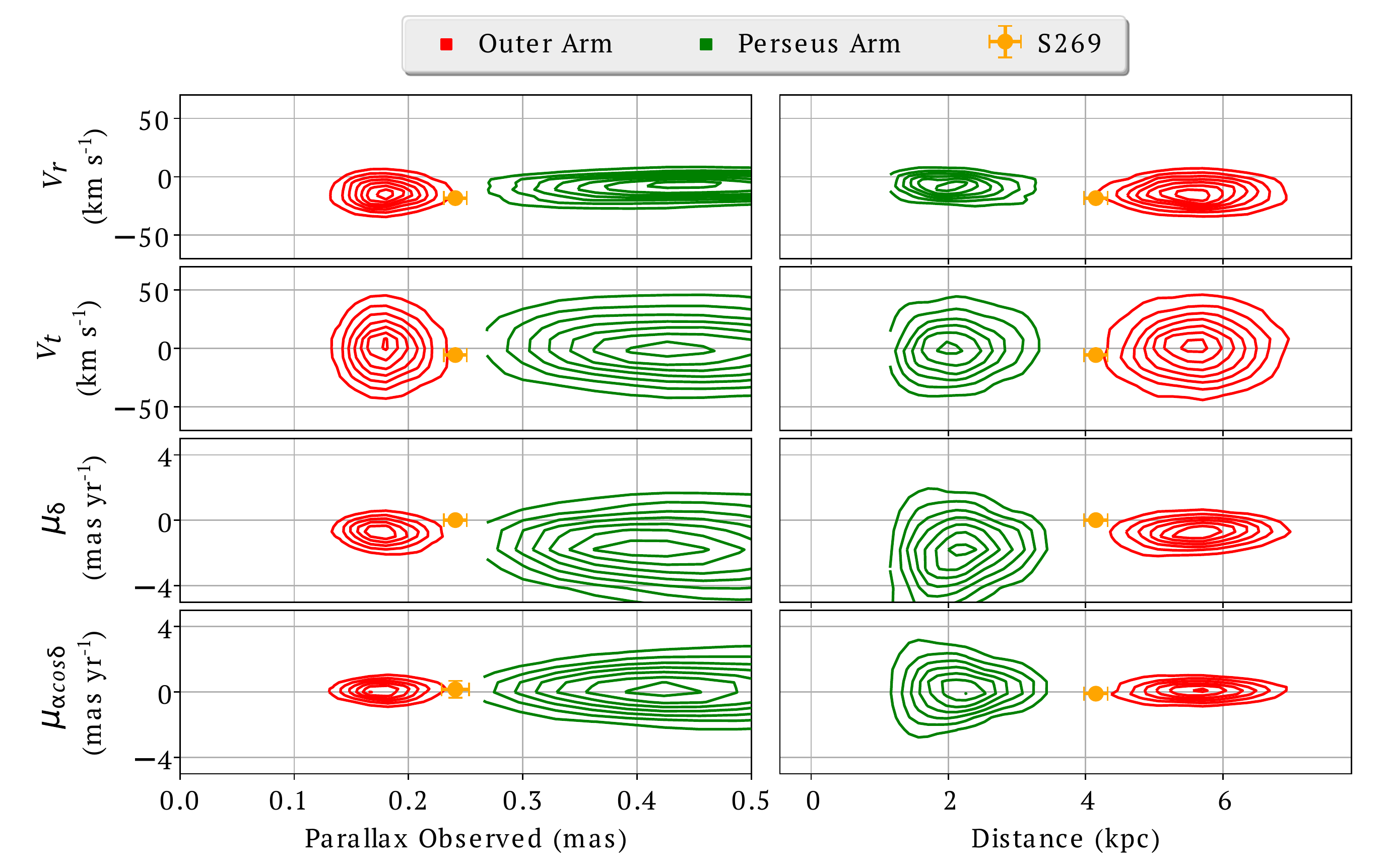}
\caption{Radial velocity ($V_{r}$), transversal velocity ($V_{t}$) and proper motion for $\rm{S\,269}$ seen from the Sun as a function of observed parallax (left plots) and distance (right plots). The contours show the distribution of sources per arm for a simulated Galactic maser distribution around the position of $\rm{S\,269}$ where the pitch angles of each arm were constant (see Sect.~\ref{s_membership_arm}). The simulations were made using the code developed by~\cite{Quiroga-Nunez2017}. Each single contour represents a percentage of the arm sources contained in a specific region of the plot from the inside as [0.14,0.28,0.42,0.56,0.70,0.84,0.98].}
\label{p_velodistri_simu}%
\end{figure*}

\subsubsection{Distance}
The combined fit of the four 22 GHz water maser spots presented in Table~\ref{t_spot_pi_ppm_fit} yielded a parallax value of $0.241 \pm 0.012$ mas, which corresponds to a distance of $4.15_{-0.20}^{+0.22}$ kpc from the Sun and $12.36 \pm 0.27$ kpc from the Galactic center~\citep[adopting $R_{\odot}=8.34$ kpc,][]{2014ApJ...783..130R}. The annual parallax is in agreement with $0.247 \pm 0.034$ mas obtained by~\cite{Asaki2014a} for the VERA data taken between 2004 and 2005.

Although~\cite{Honma2007a} reported a smaller annual parallax of $0.189 \pm 0.008$ mas, and hence a larger distance of $5.28_{-0.22}^{+0.24}$ kpc for the elongated maser spot, \cite{Asaki2014a} claimed that VERA data for that single spot yielded an inexact parallax estimate. They suggested that this problematic morphology caused an erratic positional estimate. Indeed, VERA baselines are short and few compared to the VLBA, and therefore they could not resolve and fit the inner core of the elongated spot. However, with the new VLBA observations, we have been able to fit and constrain the annual parallax to the compact core ($0.244 \pm 0.012$ mas) with respect to both extragalactic sources (see Table~\ref{t_spot_pi_ppm_fit}). This fact could explain the distance discrepancy between~\cite{Honma2007a} and this work's measurement. Also, as it was mentioned by~\cite{Asaki2014a}, parallax uncertainties reported by~\cite{Honma2007a} might be larger than quoted as the possibility of correlated positional variations among the three spots used was not considered.

\subsubsection{Peculiar Velocity}

We transformed the estimated 3D average motion of the maser spots (see Sect.~\ref{s_results_astrometric}), that is $\rm{\mu_{{\alpha}}\ cos \delta=0.16 \pm 0.26 \ mas \ yr^{-1}}$, $\rm{\mu_{{\delta}}=-0.51 \pm 0.26 \ mas \ yr^{-1}}$ and $V_\mathrm{LSR}=\rm{19.6 \pm 5 \ km \ s^{-1}}$, to the $(U,V,W)$ reference frame that rotates with the Galactic disk, yielding $U_{\rm{S\,269}}= 3 \pm 5$, $V_{\rm{S\,269}}= -1 \pm 5$ and $W_{\rm{S\,269}}= 6 \pm 5$ in $\rm{km \ s^{-1}}$, where $U$ increases toward the Galactic center, $V$ in the direction of Galactic rotation and $W$ toward the north Galactic pole. We assumed a rotation model defined by~\cite{2014ApJ...783..130R} with $R_{0}=$ 8.31 kpc and $\Theta_{0}$ = 241 km s$^{-1}$, $U_{\odot}$ = 10.5 km s$^{-1}$, $V_{\odot}$ = 14.4 km s$^{-1}$, $W_{\odot}$ = 8.9 km s$^{-1}$ and $\rm{d\Theta/dR =-0.2 \ km \ s^{-1} \ kpc^{-1}}$. The obtained values are consistent with previous findings of near-zero peculiar motion for water masers associated with HMSFRs~\citep{2014ApJ...783..130R}.

The tangential motion of $\rm{S\,269}$ allows us to constrain the Galactic rotation at 12.4 kpc radius from the center of the Milky Way. The errors in $V_{\rm{S\,269}}$ reported in this work are comparable to those reported by~\cite{Honma2007a}, where a different model for the Galactic rotation was used. As a consequence, we find that the $\rm{S\,269}$ tangential motion is within 2$\%$ of a flat Galactic rotation curve, as it was initially claimed by~\cite{Honma2007a}, albeit at a larger distance compared with this work.

\subsection{Membership in the Perseus or outer arm}
\label{s_membership_arm}

\begin{figure*}
\centering
\includegraphics[width=11.0cm]{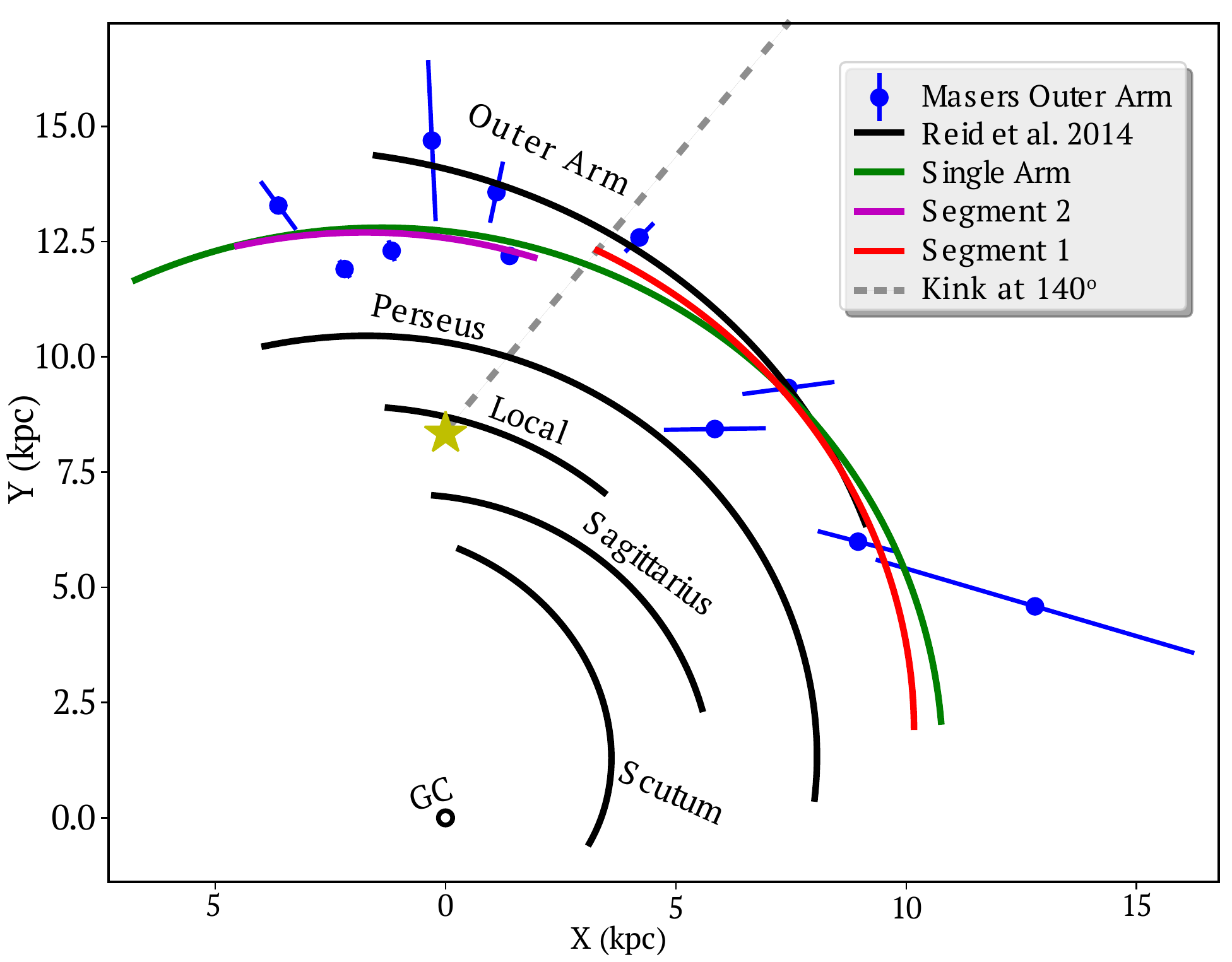}
\caption{Plan view of Galactic spiral structure. The spiral structure estimated by~\cite{2014ApJ...783..130R} is shown as black lines for reference. Maser emission from 11 HMSFRs has been used to estimate the position of the outer arm. The different outer arm descriptions discussed in Sect.~\ref{s_membership_arm} are highlighted in color curves. The Galactic center is located at (0,0), and the yellow star corresponds to the Solar position~\citep{2014ApJ...783..130R}, while the orange line demarcates the latitude of the outer arm kink suggested (i.e., 140$\degr$).}
\label{p_kinarm}%
\end{figure*}

\begin{table*}[tbp]
\caption{Astrometric information for 11 HMSFRs obtained with the VLBA in the outer arm region. These sources were used for the outer arm fitting (Sect.~\ref{s_membership_arm}).\label{t_outarm_sources}}
\begin{center}
\begin{tabular}{lllcc}
\hline \hline
\multicolumn{1}{c}{Name} & \multicolumn{1}{c}{$\alpha$}& \multicolumn{1}{c}{$\delta$} & $\pi$  & Ref. \\
    & \multicolumn{1}{c}{(hh:mm:ss)} & \multicolumn{1}{c}{($^\circ$ : $^\prime$ : $^{\prime\prime}$ )} & (mas) &  \\
\hline
G073.65+00.19 & 20:16:21.932  & +35:36:06.094 & $0.075\pm0.020$ & * \\    

G075.30+01.32 & 20:16:16.012 & +37:35:45.810 & $0.108\pm0.010$ & 1 \\
G090.92+01.48 & 21:09:12.969 & +50:01:03.664   & $0.171\pm0.031$ & * \\
G097.53+03.18 & 21:32:12.434 & +55:53:49.689   & $0.133\pm0.017$ & 2  \\
G135.27+02.79 & 02:43:28.568   & +62:57:08.388 & $0.167\pm0.011$ & 7 \\
G160.14+03.16 & 05:01:40.244 & +47:07:19.026   & $0.244\pm0.006$   &  * \\
G168.06+00.82 & 05:17:13.744 & +39:22:19.915   & $0.187\pm0.022$   &  2,3 \\
G182.67$-$03.26 & 05:39:28.425 & +24:56:31.946   & $0.157\pm0.042$   &  2,4 \\
G196.45$-$01.68 $\rm{(S\,269)}$ & 06:14:37.641  & +13:49:36.693 & $0.242\pm0.011$   & 5 \\
G211.60+01.06 & 06:52:45.321 & +01:40:23.072   & $0.239\pm0.010$ & * \\
V838$\,$Mon  & 07:04:04.822  & $-$03:50:50.636 & $0.163\pm0.016$ & 6 \\
\hline
\end{tabular}
\end{center}
{\footnotesize {\bf Notes.} The names include the galactic coordinates except for V838$\,$Mon which is (217.80,+01.05). The parallax of each source might differ from the published values in the references as we combined independent measurements (one per reference) to increase their accuracy. The parallaxes marked with * will be published as part of the BeSSeL survey (Reid priv. comm.). {\bf References:}
(1)~\cite{Sanna2012}, (2)~\cite{Hachisuka+apj17}, (3)~\cite{Honma2011}, (4) Data reanalyzed of~\cite{Hachisuka+apj17}, (5) Variance averaged between~\cite{Asaki2014a} and results of Table~\ref{t_spot_pi_ppm_fit} (6)~\cite{Sparks2008}, (7)~\cite{Hachisuka2009}.}
\end{table*}

In order to investigate whether $\rm{S\,269}$ lies within a spiral arm, we generated simulations of Galactic maser sources following the model proposed by~\cite{Quiroga-Nunez2017}. Although that model was initially developed for methanol masers associated with HMSFRs, it can be used to estimate the kinematics of other masers at certain regions of the Galaxy. There are three differences with respect to the model that~\cite{Quiroga-Nunez2017} implemented. First, we did not consider any luminosity function for water masers, since it is not necessary for our kinematic study. Second, we populated the phase space of our model with many more (up to half million) sources to allow an accurate sampling. Third, the spiral structure model follows the arm description derived by~\cite{2014ApJ...783..130R}. Although this spiral structure model was obtained using the the S269's distance estimated by~\cite{Honma2007a}, S269 was not the only source used for the spiral structure model, and also, this model was a smooth extension of the spiral arm from the first to the second quadrant. Figure~\ref{p_new_Arms} shows the simulated distribution that was obtained by this way but displaying only 2,000 sources for plotting purposes.

The phase-space density distributions of masers simulated for the outer Galaxy is shown in Fig.~\ref{p_velodistri_simu},  with the observational values of $\rm{S\,269}$ plotted. For the distributions, all the variables are measured from the Sun and simulated errors in these observables were also generated~\citep[see][for details]{Quiroga-Nunez2017}. In all cases, the kinematic parameters of $\rm{S\,269}$ seem to suggest that this source is more likely member of the outer arm than the Perseus arm, as was pointed out by~\cite{Sakai2012} based on the distance estimate made by~\cite{Honma2007a}.

\subsection{Outer arm structure}

Previous pitch angle estimates for the outer arm~\citep[e.g.,][]{2014ApJ...783..130R} were obtained based on the large distance to $\rm{S\,269}$ published by~\cite{Honma2007a}. Moreover, several sources were excluded from the pitch angle fit since they were considered interarm sources~\citep{2014ApJ...783..130R,Hachisuka+apj17}. To investigate this, we recalculated the outer arm position using recent astrometric information from ten other HMSFRs coming from the BeSSeL survey (Reid priv. comm.). These sources seem to belong to the outer arm based on their kinematics and parallaxes (see Table~\ref{t_outarm_sources}). We assess three possible scenarios for the outer arm: a single arm with a constant pitch angle, an arm with two segments that form a kink where they join, and an arm that bifurcates. In all cases, the new fit locates the outer arm in the third quadrant closer to the Sun, compared to what was previously reported~\citep{Sanna2012,2014ApJ...783..130R}.

\subsubsection{A single arm}

Following the procedure in~\cite{Reid+Apj09a} and assuming a width of $0.63 \pm 0.18$ kpc for the outer arm~\citep[estimated by][]{2014ApJ...783..130R}, we fitted 11 sources (see Table~\ref{t_outarm_sources}), finding that the spiral arm can be described using the form:

\begin{equation}
ln(R) = (2.50 \pm 0.02 ) - (\pi/180) \ (\beta - 17\fdg9) \tan(\Psi) ,
\end{equation}

\noindent where $R$ is the Galactocentric radii in kpc at a Galactocentric azimuth $\beta$ (which is zero toward the Sun and increases with Galactic longitude) and $\Psi$ the pitch angle with a value of $6\fdg2 \pm 3\fdg1$. This description applies for $ 73\degr \lesssim l \lesssim 218\degr$, which corresponds to the Galactic longitude range of the sources used.

Figure~\ref{p_kinarm} shows a plan view of the Milky Way, where the spiral arm positions estimated by~\cite{2014ApJ...783..130R} are shown as black curves for reference. The pitch angle for the outer arm calculated by~\cite{2014ApJ...783..130R} (i.e. $13\fdg8 \pm 3\fdg3$) is within the errors compared to other published values based on masers associated with massive young objects (e.g., \cite{Sanna2012} and~\cite{Hachisuka+apj17} reported $12\fdg1 \pm 4\fdg2$ and $14\fdg9 \pm 2\fdg7$, respectively). In contrast, the outer arm position with our estimate of the pitch angle (i.e., $6\fdg2 \pm 3\fdg1$) is shown in the same figure as a green line. This pitch angle is unusually small compared with previous studies ---even without considering $\rm{S\,269}$ as an outer arm source--- and entirely attributed to the sources at large Galactic longitudes ($>140\degr$) suggesting that a kink in the outer arm is another plausible explanation. Finally, although the outer arm sampling used is sparse, the reconstruction of the arm is still the best procedure with the limited astrometric solutions available.

\subsubsection{Two arm segments forming a kink}

\cite{Honig2015} analyzed the positions of a large number of HII regions in four face-on galaxies, and concluded that spiral arms seem to be composed of segments that join up and sometimes produce abrupt changes in pitch angle (kinks). We have tested if the outer arm presents a similar feature by splitting the sample into two balanced subsamples, that is five sources with $l<140\degr$ and six sources with $l>140\degr$. We estimated a pitch angle for the first segment ($l<140\degr$) $10\fdg5 \pm 5\fdg9$, and $7\fdg9 \pm 5\fdg8$ for the second segment ($l>140\degr$). The fits to both segments are shown in Fig.~\ref{p_kinarm}. While, with the small number of sources, the pitch angle estimates are quite uncertain, Fig.~\ref{p_kinarm} suggests either a kink or bifurcation in the outer arm somewhere near a longitude of $l\sim 140\degr$. Note also that this representation calls for a kink with a change of pitch angle of $\gtrsim 25\%$ ($\Delta \Psi / |\Psi|$), comparable to values of~$\sim20\%$ which are common in spiral galaxies~\citep{Savchenko2013}. Moreover, the position of the outer arm observed in HI maps by~\cite{Koo2017} for the third quadrant requires a significant displacement (or kink) within the range of 140$\degr < l < 210\degr$. Clearly, more sources with accurate measurements are needed to refine the position of a possible kink in the outer arm.

\subsubsection{Bifurcation of the arm}

As mentioned above, looking at the parallax positions of sources in Fig.~\ref{p_kinarm}, one could hypothesize that some sources follow the outer arm model of~\cite{2014ApJ...783..130R} into quadrant 3, while others rather follow the new single arm model with a smaller pitch angle or the segmented arm model, forming thus a bifurcation at l$\sim$140$\degr$. Although HI maps of the Milky Way suggest that bifurcations of the Galactic arms~\citep[e.g.,][]{Koo2017} might occur, we cannot establish if this is the case for the outer arm at the Galactic longitudes investigated here, especially in the Galactic anticenter direction, where HI maps are inaccurate due to the largest velocity component (caused by the Galactic rotation) not being radial but transversal with respect to the Sun. More sources are needed to evaluate the likelihood of this hypothesis.

\subsection{Optical members of the same stellar association}
\label{s_optical_discussion_hrd}

Massive young stars are understood to be formed from Giant Molecular Clouds that collapse generating high- and low-mass stellar cores~\citep[e.g.,][]{Tan2014}. We can search for associated stars using {\it Gaia} DR2, but given that the HMSFR that hosts the $\rm{S\,269}$~IRS~2w massive young star is located close to the Galactic plane ($b=-1\fdg46$), and at $4.15$ kpc from the Sun, only the brightest, early-type members of the same stellar association are expected to be detectable with {\it Gaia}.

We review the proper motion for the stars within 125 pc around $\rm{S\,269}$ (see Sect.~\ref{s_Gaiaresult}) using the {\it Gaia} DR2, finding that the closest ($\sim$37 pc projected distance) stellar cluster is NGC~2194. The Gaia parallax for NGC~2194 (i.e., $0.232\pm0.027$ calculated for 217 stellar members with $\sigma_{\pi}/\pi<20\%$ including zero-point correction of $-0.03$ mas) is consistent with the $\rm{S\,269}$ parallax. However, there are several reasons to suggest that $\rm{S\,269}$ may not be directly associated with NGC 2194. First, based on chemical composition,~\cite{Salaris2004} and~\cite{Netopil2016} have estimated an age of $0.87 \pm 0.19$ Gyr and $0.60 \pm 0.25$ Gyr for NGC 2194, whereas HMSFRs are expect to be two orders of magnitude younger~\citep[see, e.g.,][]{Battersby2017}. Indeed, \cite{Jiang2003} reported a dynamic age of $10^5$ yr for $\rm{S\,269}$. Second, there seems to be a serious discrepancy between the published luminosity distance~\citep[$1.9\pm0.1$ kpc,][]{Jacobson2011} and the distance estimate that one can obtain with {\it Gaia} data. 

Finally, the three closest {\it Gaia} sources to $\rm{S\,269}$~IRS~2w found within the core of the $\rm{S\,269}$ HII region defined by~\cite{Godbout1997} (i.e., $\rm{3.9 \ pc \times 2.8 \ pc}$) correspond to the three first rows in Table~\ref{t_gaia_sources}. Given that these sources have an average parallax and proper motion that are consistent with respect to the VLBA observations (i.e., $-32\pm23$ $\mu$as, $0.02\pm0.65$ and $-0.16\pm0.77$ $\rm{mas \ yr^{-1}}$), we suggest that they are likely early-type members of the same stellar association that contains $\rm{S\,269}$~IRS~2w. 
However, further studies of these companion stars and their reddening could be used to estimate the age of $\rm{S\,269}$ and possibly refine its astrometry.
\section{Conclusions}
\label{s_conclusions}

We present the results of high-accuracy VLBA observations of the $\rm{S\,269}$ region using relative astrometry. We detected nine water maser spots in $\rm{S\,269}$ that were prominent during at least three observing epochs. Four maser spots were detected in at least ten epochs, which allows a precise annual parallax fitting of $0.241 \pm 0.012$ mas corresponding to a distance of $4.15_{-0.20}^{+0.22}$ kpc.

Although the calculated distance corroborates the results previously published by~\cite{Asaki2014a}, we show that the strongest maser spot (which was left out from their analysis because of its elongated shape) yields a well-constrained annual parallax, when the inner core position is used for the fit. Also, the longevity of the elongated water maser spot in the region is remarkable as it spans more than ten years~\citep[i.e., 2004-2016, between][and our observations]{Honma2007a}, however given the significant changes in proper motion between both observational sets, we could not estimate a 10-year astrometric fit. In addition, the VLBA images and the distribution of maser spots indicate that this spot could be originated from the compression of material in a shock front that propagates perpendicular to the elongation. Moreover, water maser emission detected in the same region from 1980 to 2001 by~\cite{Lekht2001b} is likely to be the same that the VLBA detected in 2015-2016. However, the cyclic emission period previously estimated does not seem consistent with our observations.

We calculated a Galactic peculiar velocity for $\rm{S\,269}$ to be $\rm{(2\pm6}$, $\rm{4\pm14}$, $\rm{4\pm13}$) $\rm{km \ s^{-1}}$ in the ($U$,$V$,$W$) Galactic frame, which confirms that the rotation curve at large radii ($\sim$12.4 kpc) is fairly flat. On the other hand, since there is no model that ties the masers in a shock front to the motion of the underlying star, the accuracy with which we know the motion of the system is limited.

By comparing $\rm{S\,269}$'s position and proper motion with respect to other sources in the outer region of the Milky Way, we fitted the outer arm position, locating it closer to the Sun than previously thought. We explored three different scenarios: a new single outer arm pitch angle of $6\fdg2 \pm 3\fdg1$, a kink in the outer arm between two different segments and a bifurcation of the arm. Although all three are plausible explanations, the low value of a single arm pitch angle with respect to other arms and the lack of astrometric information to test a secondary segment coming from a bifurcation, lead us to
favor a kink model. This kink can be described by two segments with pitch angles of $7\fdg9 \pm 5\fdg8$ and $10\fdg5 \pm 5\fdg9$, locating the kink in the second quadrant ($\sim 140\degr$). This explanation is consistent with HI maps at $l>180\degr$, and is also supported by observations of similar features in other galaxies. Future observations are needed to assess if this is the case for the outer arm.

Finally, the {\it Gaia} DR2 catalog was inspected around $\rm{S\,269}$ for optical companions, which could be members of the stellar association. We did not find an optical counterpart for $\rm{S\,269}$~IRS~2w which could be exciting the water maser emission. However, we did find three optical sources that are likely members of the same stellar association that contains $\rm{S\,269}$~IRS~2w. Moreover, only one cluster (NGC 2194) was detected in the vicinity, but it is unlikely to be associated with $\rm{S\,269}$ given the difference in age. Future explorations of optical associations with respect to VLBI astrometric data are planned~\citep[e.g.,][]{Pihlstrom2018} to refine the criteria for optical stellar companions around HMSFRs and evolved stars.
\begin{acknowledgements}
The National Radio Astronomy Observatory is a facility of the National Science Foundation operated under cooperative agreement by Associated Universities, Inc. This work made use of the Swinburne University of Technology software correlator, developed as part of the Australian Major National Research Facilities Programme and operated under license. This work has also made use of data from the European Space Agency (ESA) mission {\it Gaia}\footnote{\url{https://www.cosmos.esa.int/gaia}}, processed by the {\it Gaia} Data Processing and Analysis Consortium (DPAC\footnote{\url{https://www.cosmos.esa.int/web/gaia/dpac/consortium}}). Funding for the DPAC has been provided by national institutions, in particular the institutions participating in the {\it Gaia} Multilateral Agreement. Moreover, this research has made use of ``Aladin sky atlas'' developed at CDS\footnote{\url{http://cds.u-strasbg.fr/}}, Strasbourg Observatory, France. The Digitized Sky Survey was produced at the Space Telescope Science Institute under U.S. Government grant NAG W-2166. The images of these surveys are based on photographic data obtained using the Oschin Schmidt Telescope on Palomar Mountain and the UK Schmidt Telescope. The plates were processed into the present compressed digital form with the permission of these institutions. This publication makes use of data products from the Two Micron All Sky Survey, which is a joint project of the University of Massachusetts and the Infrared Processing and Analysis Center/California Institute of Technology, funded by the National Aeronautics and Space Administration and the National Science Foundation. The authors sincerely acknowledge the anonymous referee for the suggestions that have improved this manuscript. L.H.Q.-N. would also deeply thank Dr. A.G.A Brown at Leiden Observatory for his comments and suggestions regarding the {\it Gaia} cross-match.
\end{acknowledgements}

\bibliographystyle{aa} 
\bibliography{quiroganunez_v2.bib}

\appendix
\label{s_appendix}
\section{Additional Tables}

\begin{table}
\caption{VLBA observational epochs for $\rm{S\,269}$ as part of the BR210E program during 2015-2016.}
\label{t_epochstime}
\begin{center}
\begin{tabular}{ccc}
\hline \hline
Epoch & Date & Time range (UTC)\\
 & (dd/mm/yyyy) & (hh:mm:ss - hh:mm:ss)\\
\hline
A &03/Aug/2015 & 11:55:29 - 18:57:42\\
B &29/Aug/2015 & 10:13:15 - 17:15:28\\
C &24/Sep/2015 & 08:31:02 - 15:33:14\\
D &18/Oct/2015 & 06:56:40 - 13:58:53\\
E &27/Jan/2016 & 00:19:33 - 07:19:45\\
F &07/Feb/2016 & 23:32:22 - 06:32:34\\
G &20/Feb/2016 & 22:41:15 - 05:41:27\\
H &29/Feb/2016 & 22:05:52 - 05:06:04\\
I &11/Mar/2016 & 21:22:37 - 04:22:49\\
J &20/Mar/2016 & 20:47:14 - 03:47:26\\
K &02/Apr/2016 & 19:56:07 - 02:56:19\\
L &15/Apr/2016 & 19:05:00 - 02:05:12\\
M &07/Aug/2016 & 11:36:47 - 18:36:58\\
N &01/Sep/2016 & 09:58:29 - 16:58:40\\
O &24/Sep/2016 & 08:28:03 - 15:28:14\\
P &22/Oct/2016 & 06:37:58 - 13:38:09\\
\hline
\end{tabular} \\
\end{center}
\end{table}

\begin{table*}[tbp]
\caption{Astrometric parameters, mean $g$ apparent magnitude and color given by {\it Gaia} DR2 for ten optical sources found in the vicinity of $\rm{S\,269}$~\citep[$10^\prime \times 7^\prime$,][]{Carpenter1990} or $\rm{12.1 \ pc \times 8.4 \ pc}$ assuming our distance estimate of 4.15 kpc. The sky distribution of the sources is shown in optical and NIR images in Fig.~\ref{p_Gaiacounter_S269_dss}. The projected distance ($\rm{d_{S\,269}}$) between the Gaia sources and maser position S269 (06:14:37.6410,+13:49:36.6930).}
\label{t_gaia_sources}
\resizebox{\hsize}{!}{
\begin{tabular}{cccccrcc}
\hline
\hline
{\it Gaia} & $\alpha$ (J2015.5) & $\delta$ (J2015.5) & $\rm{d_{S\,269}}$ & Parallax &$\rm{\mu_{\alpha} \ cos \delta}$& $\rm{\mu_{\delta}}$ \\
source ID & (hh:mm:ss) & ($^\circ$ : $^\prime$ : $^{\prime\prime}$) & (arcmin) & (mas) & (mas $\rm{yr^{-1}}$) &(mas $\rm{yr^{-1}}$)\\
\hline
3344575631369939200 & 06:14:38.73 & +13:49:45.06 & 0.30 & $0.232\pm0.043$ & $-0.012\pm0.078$& $-0.739\pm0.069$\\
3344575627071194240 & 06:14:38.99 & +13:49:43.90 &0.35 & $0.247\pm0.028$ & $-0.113\pm0.052$& $-0.247\pm0.046$\\
3344578586307442560 & 06:14:35.35 & +13:49:43.40 & 0.57 & $0.238\pm0.042$ & $0.020\pm0.077$& $-0.491\pm0.068$\\
3344575695792364672 & 06:14:45.94 & +13:50:30.56 &2.20 & $0.304\pm0.056$ & $0.711\pm0.097$& $-2.908\pm0.088$\\
3344578959966755200 & 06:14:29.28 & +13:52:32.46 &3.56 & $0.255\pm0.029$ & $0.919\pm0.058$& $-2.379\pm0.052$\\
3344580506155003776 & 06:14:41.50 & +13:53:05.03 &3.60 & $0.284\pm0.055$ & $0.280\pm0.099$& $-1.547\pm0.089$\\
3344576937039978624 & 06:14:57.98 & +13:50:12.24 & 4.97 & $0.238\pm0.047$ & $-0.070\pm0.080$& $-0.732\pm0.073$\\
3344579784600525952 & 06:14:25.88 & +13:53:43.46 & 5.01 & $0.260\pm0.037$ & $0.257\pm0.069$& $-0.864\pm0.066$\\
3344576181125735424 & 06:15:00.10 & +13:49:20.22 & 5.46 & $0.262\pm0.034$ & $0.029\pm0.064$& $-2.753\pm0.058$\\
3344573909084277632 & 06:14:41.57 & +13:44:02.90 & 5.64 & $0.279\pm0.025$ & $-0.031\pm0.043$& $-0.854\pm0.038$\\
\hline
\end{tabular}}
{\footnotesize {\bf Notes.} The sources are organized by proximity to the water mater emission with the closest first. The three first rows are the optical stars that were found within the size of the HII region $\rm{S\,269}$ defined by~\cite{Godbout1997} as $\rm{3.9 \ pc \times 2.8 \ pc}$ and are likely to be associated with the $\rm{S\,269}$ HII region (Sect.~\ref{s_optical_discussion_hrd}).}
\end{table*}

\end{document}